# Coalescence of ZnO nanowires grown from monodispersed Au nanoparticles

Liangchen Zhu, Matthew R. Phillips, and Cuong Ton-That*

*School of Mathematical and Physical Sciences, University of Technology Sydney, PO Box 123, Broadway, NSW 2007, Australia*
\* E-mail: cuong.ton-that@uts.edu.au. Tel: +61.2.95142201

**New insights into controlling nanowire merging phenomena are demonstrated in growth of thin ZnO nanowires using monodispersed Au colloidal nanoparticles as catalyst. Both nanowire diameter and density were found to be strongly dependent on the density of Au nanoparticles. Structural analysis and spectral cathodoluminescence imaging of the *c*-plane nanowire cross-sections reveal that thin isolated nanowires growing from the Au nanoparticles begin to merge and coalesce with neighbouring nanowires to form larger nanowires when their separation reaches a threshold distance. Green luminescence, which is originated from the remnants of constituent nanowires before merging, is detected at the core of fused nanowires. The distribution of nanowire diameters and green emission were found to be strongly dependent on the density of the Au nanoparticles. The merging phenomenon is attributed to electrostatic interactions between nanowire *c*-facets during growth and well-described by a cantilever bending model.**

## Introduction

Semiconductor nanowires are pivotal for many nano-size functional devices such as photodetectors, lasers and transistors, where high surface sensitivity and dense integration are desired.[1] Nanowires with precise control over dimensions and optoelectronic properties — a prerequisite for device applications — can be achieved through the bottom-up chemical vapor deposition (CVD). This growth method with a Au seeding layer has achieved the most success in producing various types of ZnO nanostructures in large quantities, in which a catalytic Au nanostructured seeding layer rapidly adsorbs metallic vapour and promotes orientated growth of nanowires.[2, 3] The Au layer thickness has been found to exert prominent influence on the density of ZnO nanowires. Compared with catalytic layers, the use of monodispersed Au nanoparticles with uniform diameter as catalyst is attractive.[4, 5] In the vapour-liquid-solid (VLS) growth of ZnO nanowires, the metal catalyst adsorb Zn vapour to a supersaturation level and the CVD process takes place underneath the Au nanoparticle catalyst, which remains at the tip of nanowires.[6-8] Whereas in the vapour-solid (VS) growth the Au nanoparticle seeds act as nucleation sites for ZnO deposition and remain on the substrate.[6-8] The underlying assumption in these growth mechanisms is that interactions between adjacent nanowires are negligible and that the size of the metal catalyst determines the diameter of the nanowires.

In the case of VS growth, ZnO nanowires orientated along the *c*-axis of the wurtzite crystal can be described as alternating stacking planes of tetrahedrally coordinated $O^{2-}$ and $Zn^{2+}$ ions, with the growing *c*-facet being either O- or Zn-terminated surface.[9-11] These uncompensated polar surfaces carry net positive or negative charges; their electrostatic interactions are known to play a critical role in the formation of nanohelices, nanobelts and nanosprings.[12-14] For example, the electrostatic forces in nanobelts result in coaxial coiling and ZnO growth along $[10\bar{1}0]$ direction.[15] Spectral cathodoluminescence (CL) imaging is a powerful method to investigate the spatial distribution of luminescence centres in ZnO nanostructures.[16] In our current work, this technique has been applied to investigate the merging of nanowires, revealing remnants of the original crystalline structure within coalesced nanowires. In this Communication, we present a new route for modulating the diameter and density of semiconductor nanowires by controlling the density of catalytic Au nanoparticles. We focus on thin, flexible ZnO nanowires grown from 5 nm Au nanoparticles and show that their close proximity to each other leads to significant electrostatic interactions causing the nanowires to bend and coalesce to form nanowires of larger diameter. This merging phenomenon leads to both the density and structural quality of ZnO nanowires highly dependent on their integration density on the substrate.

## Experimental Methods

ZnO nanowires were grown in a horizontal tube furnace using ZnO and graphite powder with a 2:1 weight ratio as the source material heated to 950°C, as described previously.[17] Si substrates were thoroughly cleaned by ultrasonication in acetone, ethanol and deionised water, successively. The dispersion of Au colloidal nanoparticles on Si substrates was achieved by sequential electrostatic adsorption of poly-l-lysine (0.1 w/v) and 5 nm Au colloidal nanoparticles using the procedure established by other workers.[18] Immersion of poly-l-lysine coated substrates in the colloidal suspension (Cytodiagnostics, diluted to $10^{12}$ to $10^{13}$ nanoparticles/mL) for 10 mins leads to Au nanoparticles adhered onto the Si substrate. The Au colloidal solutions were sonicated for 15 mins to reduce nanoparticle agglomeration before immersion of the substrate. The surface-bound density of dispersed Au nanoparticles on the substrate was controlled by varying the particle solution concentration; the surface-bound density after washing was in the range of 39 — 196 particles/$\mu m^2$. The morphology of the ZnO nanowires was investigated on Zeiss Supra 55 VP field-emission scanning electron microscope (FESEM). Spatially resolved spectroscopic information of *individual* nanowires was obtained by spectral CL imaging, in which a complete CL spectrum was measured at each pixel of an SEM image. The CL system is equipped with the option of either a Hamamatsu S7011-1007 CCD sensor or an Ocean Optics QE6500 spectrometer. These CL measurements were performed at an acceleration voltage of 5 kV (maximum photon generation depth of 82 nm).

## Results and discussion

Fig. 1 a, c show the morphology of Au nanoparticles anchored on Si substrates using the electrostatic deposition from the solutions with the indicated Au nanoparticle concentrations. Corresponding ZnO nanowires obtained by using these templates as growth catalyst are presented in Fig. 1 b, d. As a general trend, a higher density of catalytic Au nanoparticles leads to ZnO nanowires of larger diameters. More SEM images are presented in Fig. S1 (Supporting Information) to demonstrate the influence of the Au particle concentration on the nanowire diameter. The absence of Au nanoparticles at the tip area confirms that ZnO nanowires grow via the VS mechanism.[7] All the ZnO nanowires possess a hexagonal basal *c*-facet at the tips, which has the highest surface energy driving nanowire growth.[19] XRD analysis confirms that ZnO nanowires are grown along the *c*-axis orientation (Fig. S2).

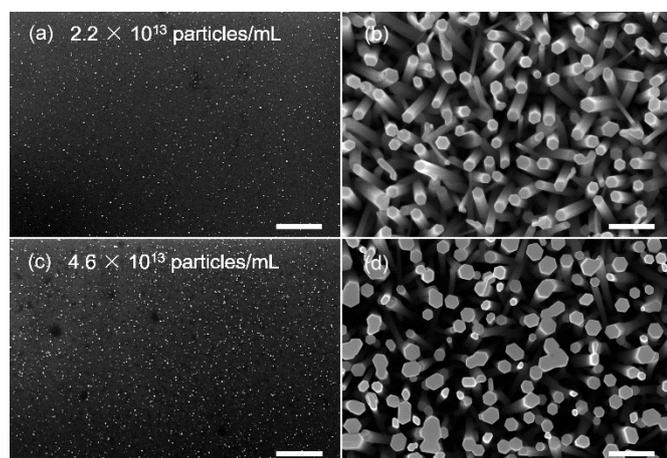

Fig. 1 (a, c) SEM images ($E_B$ = 10 keV) of catalytic 5 nm Au nanoparticles adsorbed onto charged Si substrate surfaces. The colloidal nanoparticle concentrations are displayed in the images. (b, d) ZnO nanowires grown via the VS process using the Au nanoparticle templates shown to the left. Scale bars = 500 nm.

To illustrate the relationship between the Au nanoparticle concentration and ZnO nanowire dimensions, histograms of nanowire diameters were obtained by measuring around 400 nanowires on each substrate. The nanowire diameter distribution can be fitted using a Gaussian function, which yields mean diameters of 111 ± 12 nm and 193 ± 36 nm for ZnO nanowires grown using 2.2 × $10^{13}$ and 4.6 × $10^{13}$ Au nanoparticles/mL, respectively (Fig. 2a, b). With the increase of the Au nanoparticle concentration, the mean diameter of ZnO nanowires is enlarged from 97 ± 32 to 222 ± 24 nm as the Au concentration is increased from 2.2 × $10^{13}$ to 5.6 × $10^{13}$ particles/mL, while the nanowire density drops from ~ 11 to 6 wires/$\mu m^2$, as shown in Fig. 2c. The mean nanowire diameter shows different behaviour with the increase in Au nanocatalyst concentration: below 2.2 × $10^{13}$ Au nanoparticles/mL, the nanowire diameter is statistically unchanged; while above 2.2 × $10^{13}$ Au nanoparticles/mL, the diameter of ZnO nanowires increases almost linearly with the Au concentration. The increase in the nanowire diameter indicates coalescence of thinner ZnO nanowires during growth as the Au catalytic sites on the substrate becomes denser and the nanowire separation reaches a threshold distance.

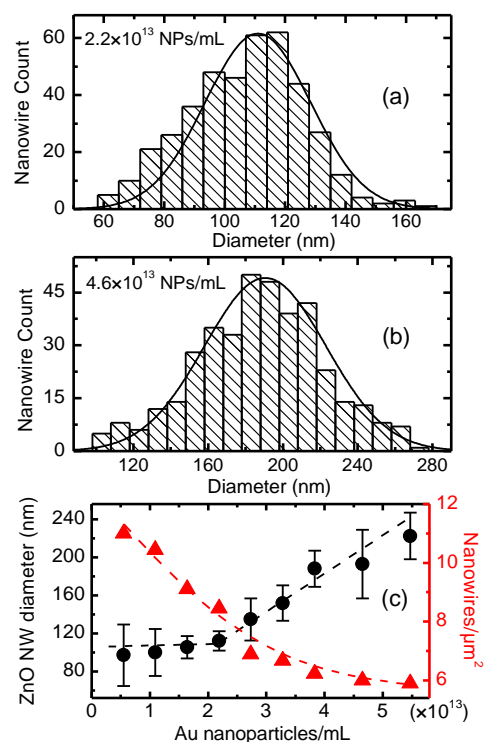

Fig. 2 (a, b) Histograms of diameters for ZnO nanowires grown from catalytic Au nanoparticles dispersed on Si substrates. The colloidal Au concentrations are indicated in the histograms. The curves represent fitted Gaussian distribution, which yields mean diameters of 111 ± 12 nm and 193 ± 36 nm. (c) ZnO nanowire diameter and nanowire density as a function of Au colloidal concentration. The black circles and the red triangles represent the diameter and density of ZnO nanowires, respectively. The error bars indicate the standard deviation.

To investigate the role of Au nanocatalysts in the growth of ZnO nanowires, Au nanoparticles were deposited through a shallow mask, producing a patterned layer of Au. This patterned template was subsequently used to grow ZnO nanowires. No ZnO nanowires were found in the Au-free regions of the template, as shown in Fig. 3a, indicating that Au catalytic nanoparticles are a prerequisite for ZnO growth. The bending of ZnO nanowires are more clearly observed in the SEM image of truncated nanowires (Fig. 3b) after ultrasonication in isopropanol, which broke the nanowires close to the merging intersection. SEM observations show that many nanowires of length greater than ~ 400 nm are bent and attracted to each other at their tips to form bundles of various numbers of nanowires. Close examination of SEM images for cleaved samples reveals some nanowires do not merge with adjacent ones but bend further away to contact with distant wires, indicating the existence of both attractive and repulsive forces.

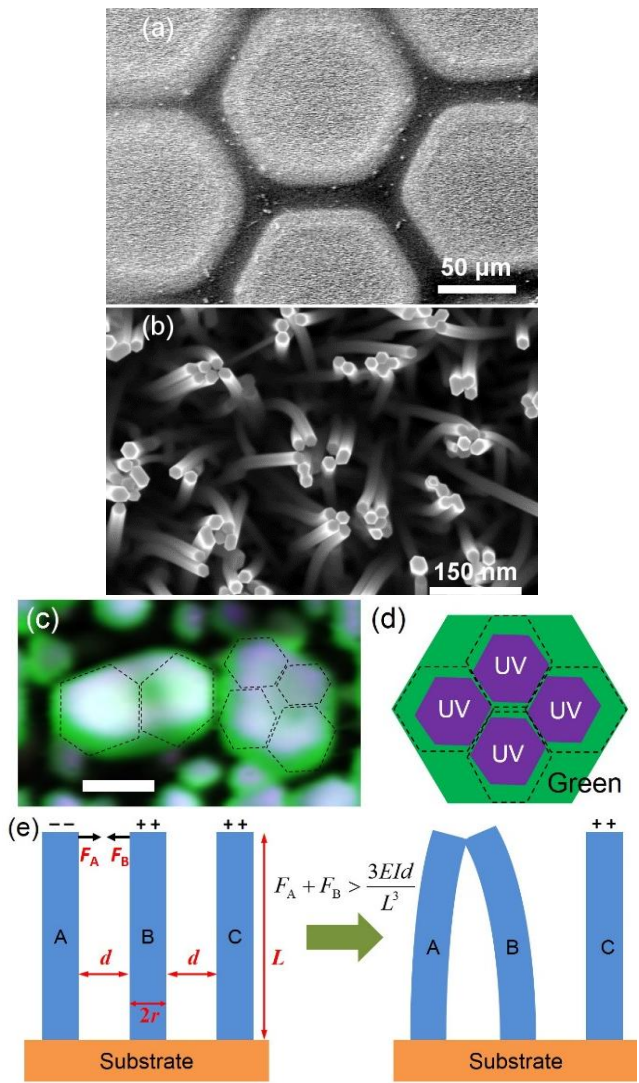

Fig. 3 (a) ZnO nanowires grown from a patterned Au layer. No ZnO nanowires were grown on the sites without Au, indicating the Au catalytic nanoparticles are pivotal to the VS-growth of ZnO nanowires. (b) SEM image showing nanowires to merge and collesce after reaching a critical length. (c) CL spectral image of fused ZnO nanowires that originates from several smaller incipient nanowires (scale bar = 200 nm). The core of the coalesced nanowires contains multi-crystalline structure with enhanced sidewall-related green emission remaining inside these nanowires. (d) Illustration of strong ultraviolet (UV) luminescence at the core of smaller individual nanowires before merging and remnant green luminescence associated with the nanowire sidewalls. (e) Schematic of nanowire bending model due to Coulomb forces between the growing $c$-facets of VS-grown ZnO nanowires.

Polar surfaces of many ionic crystals are unstable and undergo reconstruction to minimise surface energy. ZnO is an exceptional semiconductor, where both clean $(0001)$-Zn terminated and $(000\bar{1})$-O terminated surfaces have been shown to be remarkably stable.[20,21] The bending of VS-grown ZnO nanowires during growth can be attributed to the polar $c$-facets of ZnO nanowire tips, which induce long-range Coulomb interactions between the nanowires. This electrostatic interaction mechanism is consistent with the absence of coalesced tips in VLS-grown ZnO nanowires as no strong Coulomb force exists between nanowires when Au droplets sit on top of and cover the polar faces of nanowires.

The CL spectral image of the cross-sections of coalesced ZnO nanowires is shown in Fig. 3c with the schematic illustration displayed in Fig. 3d. This CL image is colour-coded to indicate regions of strong green ($h\nu$ = 2.47 eV) and ultra-violet (3.27 eV) emissions in the merging region, where the bending nanowires start to coalesce. (See Fig. S3 for the CL spectra of the nanowires.) The nanowires in Fig. 3c consist of multiple constituent nanowires (highlighted by the dashed hexagons), which are clearly distinguishable by the remnant sidewall-associated green emission that remains inside these large coalesced nanowires. We have shown previously that the green emission arises primarily from the $a$-plane sidewalls of the nanowires, while the UV luminescence dominates the nanowire core.[22] The remnants of the constituent nanowires were not completely annealed out during the short growth process. The green boundary of the coalesced nanowire is significantly larger than the remnant green region between the constituent nanowires as the crystallisation of the core takes place, driven by lower system free energy.

The density of coalesced ZnO nanowires is dependent on the concentration of the Au nanocatalysts, which determines the mean nanowire separation, $d$. When ZnO nanowires are far apart from each other, the electrostatic force between wires is negligible and the average nanowire diameter remains constant and is independent of Au colloidal concentration (Fig. 2c). When ZnO nanowires are close together, however, the Coulomb force between the polar end faces increases rapidly as ~ $1/d^2$. Nanowire tips with the same polarity repel each other while those carrying opposite charges start to attract each other and coalesce to form a single fused nanowire with a larger diameter. The increased Au nanoparticle density on the substrate reduces the average distance between individual incipient nanowires, decreasing the number of nanowires per unit area as more wires merge and coalesce. Thin nanowires have smaller moment of inertia and their mechanical stiffness can be easily overcome by the Coulomb forces between nanowires. While with thicker nanowires the mechanical force needs to be overcome surpasses the Coulomb force, so no bending and coalescence occur. To elucidate the nanowire merging, we consider the surface-charged ZnO nanowire as a cantilever with an electrostatic force exerted on the free end. The geometrical parameters (nanowire radius $r$, separation $d$, length $L$) and electrostatic forces used to describe the three-nanowire bending model are depicted in Fig. 3e. In this model, the tip of nanowire A possesses a negative charge surface while nanowires B and C carry positive charges. Suppose a force $F$ is acting on the free end of a nanowire cantilever, according to the Euler-Bernoulli beam theory,[23] the deflection at the end of the cantilever, $w$, can be shown as,

$$w = \frac{FL^3}{3EI} \quad (1)$$

where $E$ is the elastic modulus of ZnO and $I$ is the moment of inertia for a hexagonal beam, $I = (5\sqrt{3}/16)^4$. The bending and coalescence of ZnO nanowires occur at specific stages of the growth when nanowires reach an appropriate length. For nanowires A and B to make contact, their deflection at the free end must be larger than their separation, $d$:

$$w_A + w_B = \frac{F_A L^3}{3EI} + \frac{F_B L^3}{3EI} > d \quad (2)$$

Given the high aspect ratios of the nanowires, the tips can be modelled as a point charge, $Q = (3\sqrt{3}/2)r^2 \cdot \sigma$, where $\sigma$ is the surface charge density. Hence, the electrostatic forces acting on the free end of nanowires A and B can ultimately be expressed as:

$$F_A = k\frac{Q^2}{d^2} + k\frac{Q^2}{(2d)^2} = k\frac{135 r^4 \cdot \sigma^2}{16 d^2} \quad (3a)$$

$$F_B = k\frac{Q^2}{d^2} + k\frac{Q^2}{d^2} = k\frac{27 r^4 \cdot \sigma^2}{2 d^2} \quad (3b)$$

Substituting $F_A$ and $F_B$ into (2) gives an estimate of the nanowire surface charge required to bend nanowires:

$$\sigma > \frac{d}{3L}\sqrt{\frac{5\sqrt{3}Ed}{13kL}} \quad (4)$$

For the ZnO nanowire system shown in Fig. 3 with average wire length $L = 400$ nm, distance $d = 50$ nm, and ZnO elastic modulus $E = 29$ GPa,[24] the minimum charge density required is $\sigma_{min} = 0.022$ C/m$^2$. This value is of the same order of magnitude as the spontaneous polarization of bulk ZnO obtained from *ab initio* calculations (0.06 C/m$^2$).[25] Our calculation based on this qualitative model thus supports the electrostatic interaction mechanism that drives the bending of nanowires.

An important aspect of ZnO nanowire growth in this approach is the dependence of defect formation on the density of Au catalytic nanoparticles. Fig. S3 shows that CL spectra of the nanowires, which exhibit two main luminescence features: a sharp near-band-edge (NBE) emission at 3.27 eV and a symmetrical green luminescence (GL) peak at 2.47 eV, which has been proved to be related to oxygen vacancy ($V_O$) defects.[26] To investigate the role of catalytic Au, the relative integrated intensity of the $V_O$-related GL is plotted as a function of Au nanoparticle concentration (inset of Fig. S3). The intensity ratio $I_{GL}/I_{NBE}$ rises monotonously from 4.8 to 7.4 over the Au concentration range used in this study, indicating an increase in $V_O$ defect concentration. This defect density dependence on nanowire diameter can be explained by the annihilation kinetics of defects during high-temperature growth, in which $V_O$ defects in small ZnO nanowires are annealed out more readily than in large nanowires.

## Conclusions

In conclusion, ZnO nanowires have been successfully grown via the VS process by using Au catalytic nanoparticles. The width of the ZnO nanowires becomes larger as the density of Au nanoparticles increases. Some remnant of green luminescence from the *a*-plane sidewalls of constituent nanowires were found within the coalesced nanowire. A nanowire bending model is established by investigating the Coulomb forces between the *c*-facets of nanowires. Optical characterisation shows that larger nanowires contain more oxygen vacancies than smaller ones. ZnO nanowires of different diameters can be grown using colloidal Au nanoparticles by simply controlling the Au nanoparticle concentration, which can provide a new route for modulating the diameter of semiconductor nanowires.

## Acknowledgements

This work was supported by Australian Research Council Discovery Grant DP0986951. We thank Geoff McCredie and Mark Berkahn for technical assistance with the growth of nanowires and XRD measurements.

## Notes and references